\documentclass[%
 aip,
 jmp,%
 amsmath,amssymb,
 reprint,%
]{revtex4-1}

\usepackage{graphicx}
\usepackage{dcolumn}
\usepackage{bm}
\usepackage{amssymb}   
\usepackage{times}
\usepackage{color}

\begin{document}

\preprint{AIP/123-QED}

\title{High-efficiency waveguide couplers via impedance-tunable transformation optics}

\author{Jun~Cao}
\affiliation{National Research Center for Optical Sensing/Communications Integrated Networking, Department of Electronics Engineering, Southeast University, Nanjing 210096, China }
\affiliation{Department of
Physics,
Nanjing Xiaozhuang University, Nanjing 211171,  China}
\author{Lifa~Zhang}
\affiliation{Department of Physics, The University of Texas at Austin, Austin, Texas 78712, USA}
\author{Weifeng~Jiang}
\affiliation{National Research Center for Optical Sensing/Communications Integrated Networking, Department of Electronics Engineering, Southeast University, Nanjing 210096, China }
\author{Senlin~Yan}
\affiliation{Department of
Physics,
Nanjing Xiaozhuang University, Nanjing 211171,  China}
\author{Xiaohan~Sun}
\email{xhsun@seu.edu.cn}
\affiliation{National Research Center for Optical Sensing/Communications Integrated Networking, Department of Electronics Engineering, Southeast University, Nanjing 210096, China }

\date{28 July 2014}

\begin{abstract}
We design compact waveguide couplers via impedance-tunable transformation optics. By tuning impedance coefficients in the original space, two-dimensional metallic and dielectric waveguide couplers are designed with a high efficiency. Through tuning refractive index simultaneously, we find that the transformation medium inside a designed metallic waveguide coupler can be a reduced-parameter material for coupling waves between waveguides with arbitrary different cross sections and embedded media. In the design of dielectric waveguide couplers, we apply two different schemes: one is that both core region and its cladding region are contained in a transformed space, the other is only core region contained in the transformed space. The former has a very high efficiency near $100\%$; the latter is less efficient with a very small decline which can be a simplified candidate in the design of near-perfect dielectric waveguide couplers. The transformation medium for dielectric waveguide couplers can also be reduced-parameter material by selecting appropriate refractive index coefficients. Two-dimensional numerical simulations confirm our design with good performances.
\end{abstract}

\maketitle

Couplers are very essential electromagnetic (EM) devices, which act as intermediary components to reduce mode mismatch in order to transfer light efficiently between waveguides of different cross sections and embedded media. The conventional method to reduce the coupling loss is to introduce a connected taper with a gradient change of size\cite{Almeida03,Shiraishi07,Sun09}; however, it is very space consuming. With the progress of integrated optical circuits, the fiber-to-chip coupling remains a critical problem. To solve it, various couplers via different mechanisms have been proposed, such as the grating coupler\cite{Taillaert06,Roelkens08,Schmid09,Liu10}, parabolic reflector\cite{Dillon08}, and luneburg lens\cite{Arigong13}; but the reported coupling efficiency are still very low. Thus to efficiently couple EM waves in waveguides is still a challenging topic up to now.

Transformation optics, as a unconventional theory reported by Pendry \cite{Pendry06} and Leonhardt \cite{Leonhardt06}, provides a powerful tool to design various optical devices \cite{Schurig06,Huangfu08,Lai09,Chen07,Rahm08a,
Yang08,Jiang08a,Rahm08b}. However, in the technique applied in some cases\cite{Rahm08c,Jiang08b,Kwon09} to manipulate the intensity of the electromagnetic fields out of a transformation medium,
there is a mismatch of impedance at boundaries and the resulted reflections limit its applications. Different taper structures placed with transformation medium have been applied to the design of waveguide couplers\cite{Tichit10}, but the impedance mismatch has not been solved thus the coupler performance are not satisfied. In Ref.~\cite{Xu11}, impedance has been matched by inserting another special impedance-matched media in one of waveguides, however it can not be easily generalized to arbitrary coupling waves. Very recently, we proposed a generalized theory of impedance-tunable transformation optics in the geometric optics limit \cite{Cao14}, which not only can manipulate impedance match but also can provide more flexibilities to select realizable transformation medium.

In this letter, we design two-dimensional (2D) metallic and dielectric waveguide couplers using the impedance-tunable transformation optics. By tuning the impedance coefficient of the coupler, the mode waves can be coupled efficiently between two waveguides with arbitrarily different cross sections and embedded media. Through tuning the refractive index in the original space, the transformation medium can be reduced-parameter materials (magnetic-response-only materials for TE polarization or dielectric-response-only materials for TM polarization). Different schemes have been implemented and compared for the design of high-efficiency dielectric waveguide couplers.

In the theory of impedance-tunable transformation optics\cite{Cao14},
the relative permittivity $\varepsilon^{i^\prime j^\prime}$ and permeability $\mu^{i^\prime j^\prime}$ of the transformation medium for a given coordinate transformation $x^\prime=x^\prime(x)$ can be expressed as
\begin{eqnarray} \label{eq1}
\varepsilon^{i^\prime
j^\prime}=|det(A_i^{i^\prime})|^{-1}A_i^{i^\prime}A_j^{j^\prime}\varepsilon\delta^{ij}/k,
\nonumber\\
 \mu^{i^\prime
j^\prime}=|det(A_i^{i^\prime})|^{-1}A_i^{i^\prime}A_j^{j^\prime}k\mu\delta^{ij}.
\end{eqnarray}
Here $A_i^{i^\prime}=\frac{\partial(x^\prime,y^\prime,z^\prime)}{\partial(x,y,z)}$
denotes a Jacobian tensor between the transformed space
$(x^\prime,y^\prime,z^\prime)$ and the original space $(x,y,z)$, and the impedance coefficient $k$ is a spatial continuous function.

\begin{figure}[t]
\includegraphics[width=2.6 in,  angle=0]{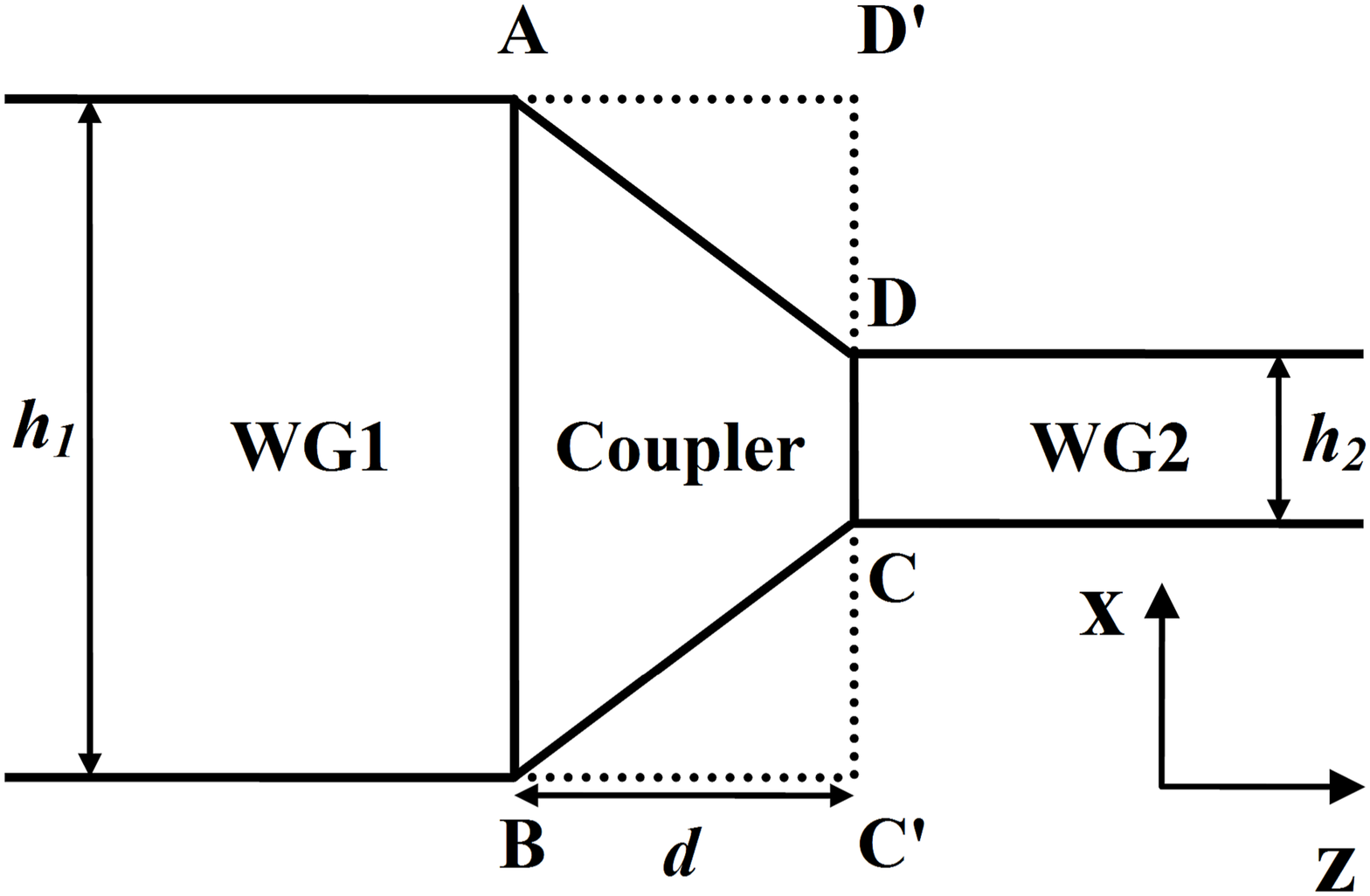}
\caption{ \label{fig1} (Color online)   Schematic diagram of the waveguide coupler using the impedance-tunable transformation optics. The width, relative permittivity, permeability of the waveguide WG1 and WG2 are $h_1$, $\varepsilon_1$,$\mu_1$ and $h_2$, $\varepsilon_2$, $\mu_2$ respectively; the length of the coupler is $d$, and embedded in a transformation medium. }
\end{figure}

Based on the strategy of impedance-tunable transformation optics, a 2D compact waveguide coupler is designed as shown in Fig.~\ref{fig1}, where waveguide WG1 and WG2 of different width $h_1$  and $h_2$ are connected by a linear taper structure ABCD of length $d$ as a waveguide coupler.  The relative permittivity and  permeability of the media in WG1 and WG2 are $\varepsilon_1$, $\mu_1$ and $\varepsilon_2$, $\mu_2$ respectively. To couple waves efficiently from WG1 to WG2 in the $+z$ direction, the region ABCD are embedded with transformation medium, which compresses the rectangular region ABC'D' in the original space to trapezoidal region ABCD in the transformed space. We apply the transformation defined in the 2D compressor/expander design\cite{Rahm08b,Cao14}:
\begin{equation}
 x' = x[1 - \frac{z}{d}(1 - \gamma )],\;\; y' = y,\; \; z' = z \\
\end{equation}
where $\gamma  = \frac{{h_2 }}{{h_1 }}$ is the compression coefficient.

Different from Ref.~\cite{Cao14} where the background of transformation media is supposed to be air, arbitrary background media $\varepsilon_1$, $\mu_1$ and $\varepsilon_2$, $\mu_2$ are discussed here. Thus we reset the material parameters in the original space to calculate a new impedance function $k$. Without changing the refractive index, the permittivity and permeability of the original medium are set to be $\varepsilon_1/k$ and $k\mu_1$. No unique function of $k$ can be selected, but there is no guarantee that the transformation medium can be a reduced-parameter material. A normal incident plane wave illuminating upon the coupler is discussed in the geometrical optics limit. Similar to the derivation in Ref.~\cite{Cao14}, the function $k$ can be set as $k = \frac{d}{{d - z'(1 - \gamma \sqrt {\frac{{\varepsilon _2 \mu _1 }}{{\varepsilon _1 \mu _2 }}} )}}$
for TE polarization, and $k = \frac{{d - z'(1 - \gamma \sqrt {\frac{{\varepsilon _1 \mu _2 }}{{\varepsilon _2 \mu _1 }}} )}}{d}$
for TM polarization. Note that
only when the impedance of embedded media in WG1 and WG2 are equal,
$\varepsilon _1 \mu _2  = \varepsilon _2 \mu _1 $,
the function $k$ are the same as that in Ref.~\cite{Cao14}, which leads to a reduced-parameter transformation medium.

\begin{figure}[t]
\includegraphics[ width=3.40 in,  angle=0]{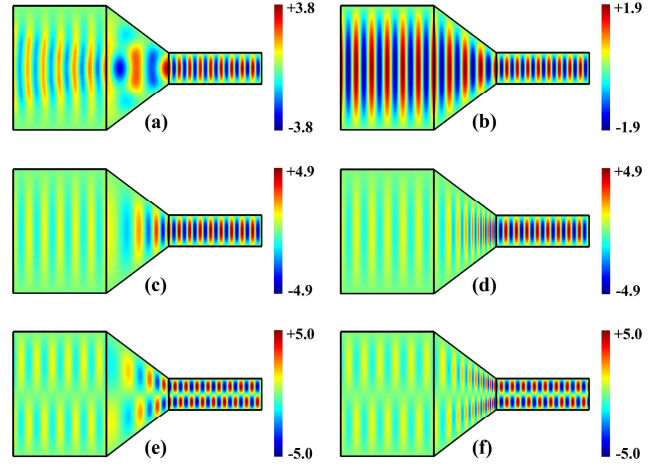}
\caption{ \label{fig2} (Color online) The normalized $y$ direction electric field distribution of a waveguide coupler. (a) A traditional case  without transformation; (b) An impedance-nontunable transformation case; (c) and (d) are impedance-tunable transformation cases for TE$_{01}$ mode with $p=1$ and $p=3$, respectively; (e) and (f) are impedance-tunable transformation cases for TE$_{02}$ mode with $p=1$ and $p=3$, respectively.}
\end{figure}

Reduced sets of material properties are generally favored in the two-dimensional transformation-optical design \cite{Schurig06,Cai07} due to its easier realization. In this paper, in order to obtain a reduced-parameter transformation medium for arbitrary conditions, we generalize the theory of impedance-tunable transformation optics to the design of  waveguide coupler that the refractive index in the original space can also be tunable, and the relative permittivity and permeability in the original space are reset to be
$\varepsilon ^{ij}  = \varepsilon(z)\delta ^{ij} /k$
and $\mu ^{ij}  = k\mu(z)\delta ^{ij} $. The refractive index
$n = \sqrt {\varepsilon(z)\mu(z)} $ in the original space is a $z$-dependent continuous function, thus it will not change the light rays compared to the conventional transformation. For TE polarization we  set $\varepsilon(z) = p$,
$\mu(z) = p\frac{{\mu _{1}  + \frac{{z'}}{d}(\mu _{2}  - \mu _{1} )}}{{\varepsilon _{1}  + \frac{{z'}}{d}(\varepsilon _{2}  - \varepsilon _{1} )}}$ and  for TM polarization $\mu(z) = p$,
$\varepsilon(z) = p\frac{{\varepsilon _{1}  + \frac{{z'}}{d}(\varepsilon _{2}  - \varepsilon _{1} )}}{{\mu _{1}  + \frac{{z'}}{d}(\mu _{2}  - \mu _{1} )}}$. Note that coefficient $p$ has not change the impedance but change its refractive index of the transformation medium, we define it as refractive index coefficient, large $p$ will extend the application of the geometric optics approximation, and improve the performance of the coupler at low frequencies.

Applying the continuity of the total tangential electric and magnetic fields, one can obtain $k = \frac{d}{{d - z'(1 - \gamma )}}$ for TE polarization and $k = \frac{{d - z'(1 - \gamma )}}{d}$ for TM polarization, which satisfy the impedance match condition at the boundaries AB and CD of the coupler.  The parameters of the transformation medium associated with TE polarization can be written as:
\begin{equation}\label{eq_te}
\begin{array}{l}
 \varepsilon _{yy}  = p \\
 \mu _{xx}  = p(1 + \frac{{(1 - \gamma )^2 d^2 x'^2 }}{{[d - z'(1 - \gamma )]^4 }})\frac{{\mu _{1}  + \frac{{z'}}{d}(\mu _{2}  - \mu _{1} )}}{{\varepsilon _{1}  + \frac{{z'}}{d}(\varepsilon _{2}  - \varepsilon _{1} )}} \\
 \mu _{xz}  = \mu _{zx}  = p\frac{{ - (1 - \gamma )d^2 x'}}{{[d - z'(1 - \gamma )]^3 }}\frac{{\mu _{1}  + \frac{{z'}}{d}(\mu _{2}  - \mu _{1} )}}{{\varepsilon _{1}  + \frac{{z'}}{d}(\varepsilon _{2}  - \varepsilon _{1} )}} \\
 \mu _{zz}  = p\frac{{d^2 }}{{[d - z'(1 - \gamma )]^2 }}\frac{{\mu _{1}  + \frac{{z'}}{d}(\mu _{2}  - \mu _{1} )}}{{\varepsilon _{1}  + \frac{{z'}}{d}(\varepsilon _{2}  - \varepsilon _{1} )}}. \\
 \end{array}
\end{equation}
The material based on the above transformation is of magnetic response only. We also can obtain a  dielectric-response-only transformation medium associated with TM polarization as:
\begin{equation}\label{eq_tm}
\begin{array}{l}
 \mu _{yy}  = p \\
 \varepsilon _{xx}  = p(1 + \frac{{(1 - \gamma )^2 d^2 x'^2 }}{{[d - z'(1 - \gamma )]^4 }})\frac{{\varepsilon _{1}  + \frac{{z'}}{d}(\varepsilon _{2}  - \varepsilon _{1} )}}{{\mu _{1}  + \frac{{z'}}{d}(\mu _{2}  - \mu _{1} )}} \\
 \varepsilon _{xz}  = \varepsilon _{zx}  = p\frac{{ - (1 - \gamma )d^2 x'}}{{[d - z'(1 - \gamma )]^3 }}\frac{{\varepsilon _{1}  + \frac{{z'}}{d}(\varepsilon _{2}  - \varepsilon _{1} )}}{{\mu _{1}  + \frac{{z'}}{d}(\mu _{2}  - \mu _{1} )}} \\
 \varepsilon _{zz}  = p\frac{{d^2 }}{{[d - z'(1 - \gamma )]^2 }}\frac{{\varepsilon _{1}  + \frac{{z'}}{d}(\varepsilon _{2}  - \varepsilon _{1} )}}{{\mu _{1}  + \frac{{z'}}{d}(\mu _{2}  - \mu _{1} )}}. \\
 \end{array}
\end{equation}
\begin{figure}[t]
\includegraphics[width=3.4 in,  angle=0]{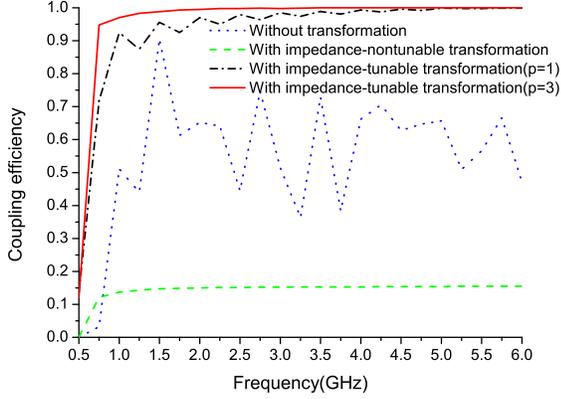}
\caption{ \label{fig3} (Color online) Coupling efficiency of a metallic waveguide coupler for TE$_{01}$ mode as a function of frequency. The blue dotted and green dashed line correspond to couplers without transformation and with
impedance-nontunable transformation, respectively; the black dash-dotted and red solid line correspond to couplers with impedance-tunable transformation for $p=1$ and $p=3$, respectively. }
\end{figure}

To investigate the performance of the waveguide coupler embedded with impedance-tunable transformation medium, we do two-dimensional numerical simulations using COMSOL Multiphysics for TE-mode waves incidence (similar calculation for TM-mode waves can also be done). For a metallic waveguide, the calculation domain is bounded by a perfectly electric conductor. In our simulations, without loss of generality, the width of the waveguide WG1 and WG2 are set to be $h_1=0.4$ m and $h_2=0.1$ m respectively, and $\varepsilon_{1} =4$, $\mu_{1} =1$, and $\varepsilon_{2} =1$, $\mu_{2} =9$ respectively. The length of the coupler is $d=0.2$ m. TE mode waves are excited at a port with an incident frequency $f=3$ GHz.  Fig.~\ref{fig2} shows electric field simulation results, in which the inset (a)-(d) are for TE$_{01}$ mode, and the inset (e)-(f) are for
TE$_{02}$ mode. Figure~\ref{fig2}(a) shows a traditional design case without transformation medium in the coupler, where the coupling efficiency $\eta  = W_2 /W_1$ is $51\%$ due to the obvious reflections ($W_1$ and $W_2$ are the coupler input and output power, respectively). For a conventional transformation medium embedded in the coupler (an impedance non-tunable case for $k=1$ ), as shown in Fig.~\ref{fig2}(b), although the wave profile are preserved well, even more serious reflections occur at the exit boundary of the coupler, resulting in an amplitude modulation of the incoming wave, and the simulated coupling efficiency is only $15\% $. While the impedance-tunable transformation medium embedded in the coupler with $p=1$, Fig.~\ref{fig2}(c) shows a good performance with the coupling efficiency near $98\% $, and improved to $99\% $ in Fig.~\ref{fig2}(d) with a larger refractive index coefficient $p=3$, which is almost reflectionless. The coupler using the impedance-tunable transformation optics can also be applied to couple high-order mode waves efficiently; Fig.~\ref{fig2}(e) and Fig.~\ref{fig2}(f) are simulated results of TE$_{02}$ mode with $p=1$ and $p=3$ respectively, with a similar coupling efficiency to that of TE$_{01}$ mode.

Figure~\ref{fig3} shows the coupling efficiency of TE$_{01}$ modes for an impedance-tunable coupler, an impedance-nontunable one and a traditional one without transformation simulated from $0.5$ GHz to $6$ GHz. As can be clearly seen from Fig.~\ref{fig3}, the traditional transformation optics does not exhibit enough competitive advantages compared to the conventional case without transformation, and thus limits its applications in many cases. However, by introducing a tunable impedance, we can obtain a coupler with extremely high efficiencies except in the vicinity of the cutoff frequency. In the whole range of frequency, the impedance-tunable coupler has a best performance compared to the impedance-nontunable one via traditional transformation optics and the traditional one without transformation. The inevitable unperfect coupling performance at low frequencies due to the geometric optics limit can be improved through setting large refractive index coefficient $p$, the cost is that the relative permittivity of the transformation medium is not 1 and extreme larger parameters will increase the difficulty of  realization of designed transformation medium. Nevertheless we still can properly increase $p$ to increase the performance at low frequencies while it is not difficult to be realized in designed medium.

\begin{figure}[t]
\includegraphics[ width=3.40 in,  angle=0]{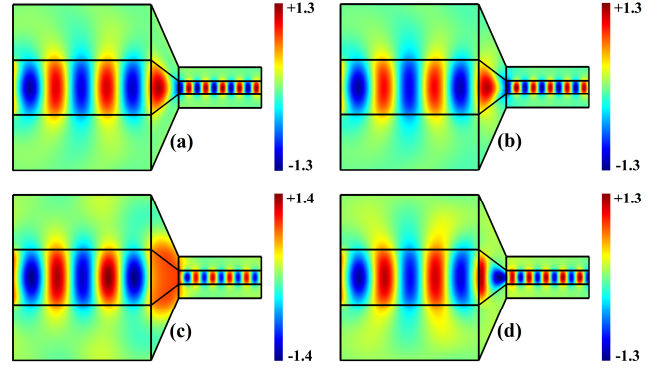}
\caption{ \label{fig4} (Color online) The normalized $y$ direction electric field distribution of a dielectric waveguide coupler with total transformation (a), with only core transformation (refractive index nontunable) (b), with only core transformation (refractive index tunable with $p=1$) (c) and with only core transformation (refractive index tunable with $p=6$) (d). }
\end{figure}

Beside a metallic waveguide, an efficient dielectric waveguide coupler is also very important in optical design. Using the impedance-tunable transformation optics method, a 2D compact dielectric waveguide coupler can be designed with high efficiency and less space occupation.

\begin{figure}[t]
\includegraphics[width=3.4 in,  angle=0]{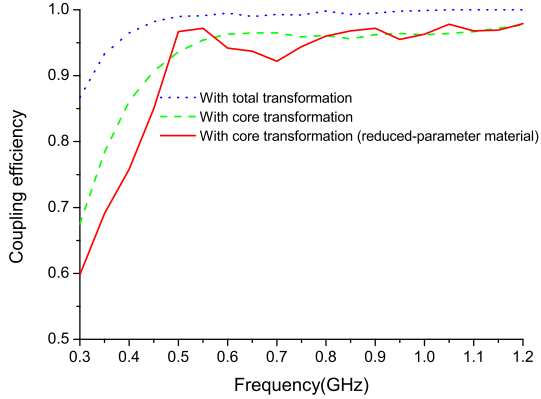}
\caption{ \label{fig5} (Color online) Coupling efficiency of a dielectric waveguide coupler as a function of frequency. The blue dotted, green dashed and red solid lines correspond to the couplers with total transformation, with only core transformation, and with only core transformation while the transformation medium is reduced-parameter material (refractive index tunable with $p=6$) , respectively.}
\end{figure}

For a dielectric waveguide, distribution of the EM field can be divided into two regions: a concentrated dielectric core and an evanescent air cladding. Therefore to couple waves with a high efficiency, the transformed space can also be divided into an inner space (major contribution) and an outer space (minor contribution). For the total transformation the embedded medium can be obtained through setting different original spaces and tuning different impedance coefficients for inner and outer parts. Note that in order to preserve the profile of the mode waves in the dielectric waveguide coupler the refractive index in the inner original space can not be tunable, resulted a non-reduced sets of material properties in the inner space of the coupler. However at high frequencies or in high-index-contrast waveguide, a simple method is acceptable to omit evanescent energy and only focus on the core energy coupling. Thus only inner transformation medium (core transformation) can be considered with a slightly reduced coupling efficiency as we will see in the simulations latter; and the transformation medium also can be reduced-parameter material if we carefully set the refractive index coefficient $p$.

In total transformation, the relative permittivity and permeability in the inner and outer original space are set to be
$\varepsilon ^{ij}  = \varepsilon_1\delta ^{ij} /k_1$
, $\mu ^{ij}  = k_1\delta ^{ij} $ and $\varepsilon ^{ij}  = \delta ^{ij} /k_2$, $\mu ^{ij}  = k_2\delta ^{ij} $ respectively.
One can obtain $k_1 = \frac{d}{{d - z'(1 - \gamma \sqrt {\frac{{\varepsilon _2 }}{{\varepsilon _1 }}} )}}$,
$k_2 = \frac{d}{{d - z'(1 - \gamma )}}$ for TE polarization,
and $k_1 = \frac{{d - z'(1 - \gamma \sqrt {\frac{{\varepsilon _1 }}{{\varepsilon _2 }}} )}}{d}$,
$k_2 = \frac{{d - z'(1 - \gamma )}}{d}$ for TM polarization.

In numerical simulations, the relative permittivity of cores in dielectric waveguide WG1 and WG2 are set as $\varepsilon_1=2$ and $\varepsilon_2=20$, the width of the core in WG1 and WG2 are 0.4 m and 0.1 m, respectively, the coupler length is 0.2 m. The incident frequency is set to be 0.6 GHz, thus only TE$_{01}$ mode can exist in WG1 and WG2. Fig.~\ref{fig4}(a) shows that TE$_{01}$ mode waves have been coupled from WG1 to WG2 with a high coupling efficiency over $99\% $. As a simplified method with only core transformation, Fig.~\ref{fig4}(b) shows a little difference compared to Fig.~\ref{fig4}(a), with a slightly reduced coupling efficiency but also near to $97\% $. Note that in simulations of Fig.~\ref{fig4}(b), the refractive index in the original core space is fixed, resulting in no reduced-parameter transformation medium. If we let the refractive index tunable, reduced-parameter material can be obtained. But such reduced parameter may make refractive index of the core less than its of cladding thus cause large radiation loss if we select a small refractive index coefficient $p$ ; as shown in Fig.~\ref{fig4}(c) the coupling efficiency is only $61\% $ with $p=1$.  While if we set a large refractive index coefficient $p=6$ in Fig.~\ref{fig4}(d), the coupling efficiency near $95\% $ can be obtained.

As a further step, the coupling efficiency with only core transformation and total transformation have been simulated at different incident frequencies from 0.3 GHz to 1.2 GHz in Fig.~\ref{fig5}. The coupling efficiency of the dielectric waveguide coupler increases at high frequencies, where high-order mode waves may propagate in the waveguide, but the performance never be affected with the coupling efficiency near to $100\% $ by total transformation design, and a very small decline of efficiency for only core transformation, which can be a proper candidate in the design of near-perfect dielectric waveguide coupler.  The transformation medium also can be reduced-parameter material by setting a appropriate tunable refractive index, but with a lower coupling efficiency.

In conclusion, to reduce the mode mismatch, the impedance-tunable transformation optics have been applied to the design of metallic and dielectric waveguide couplers with a high efficiency. It is found that reduced-parameter medium can be obtained  by tuning the refractive index in the original space. The declined performance at low frequencies can be improved by setting a large refractive index. The method can be further applied to three-dimensional waveguide coupler design in the future, especially for fiber-to-chip coupling.

{\it Acknowledgements} -- We acknowledge support from  the Frontier Research
and Development Projects of Jiangsu Province, China, under grants of BY2011147 and BY2013073.

\end{document}